\documentclass[preprint,12pt]{article}

\RequirePackage{amsthm,amsmath,amsfonts,amssymb}
\RequirePackage[numbers,sort&compress]{natbib}
\RequirePackage[colorlinks,citecolor=blue,urlcolor=blue]{hyperref}
\RequirePackage{graphicx}
\RequirePackage{bm}

\theoremstyle{plain}

\newtheorem{theorem}{Theorem}[section]

\theoremstyle{remark}

\newtheorem{remark}[theorem]{Remark}

\newcommand{\R}{\mathbb R}
\newcommand{\N}{\mathbb N}

\def\be#1\ee{\begin{equation}#1\end{equation}}
\newcommand{\fer}[1]{(\ref{#1})}

\numberwithin{equation}{section}

\newcommand{\bq}{\begin{equation}}
\newcommand{\eq}{\end{equation}}
\newenvironment{equations}{\equation\aligned}{\endaligned\endequation}
\def\bqa{\begin{eqnarray}}
\def\eqa{\end{eqnarray}}

\def\e{\epsilon}

\def\bx{{\bf x}}
\def\bX{{\bf X}}
\def\bY{{\bf Y}}
\def\bW{{\bf W}}
\def\by{{\bf y}}
\def\bm{{\bf m}}
\def\bxi{{\pmb\xi}}

\def\aa{{\bf a}}
\def\ab{{\bf b}}

\newcommand{\bd}{\begin{displaymath}}
\newcommand{\ed}{\end{displaymath}}
\newcommand{\ba}{\begin{eqnarray}}
\newcommand{\ea}{\end{eqnarray}}

\def\ff{\widehat f}

\def\gg{\widehat g}
\def\hh{\widehat h}

\def\N{\mathbb{N}}
\def\R{\mathbb{R}}

\begin{document}





\noindent \textbf{\Large Measuring multidimensional inequality: a new proposal based on the Fourier transform}
\vspace{3cm}


\noindent Paolo Giudici\\
\noindent Department of Economics and Management, University of Pavia (Italy)\\
\noindent email: paolo.giudici@unipv.it

\vspace{1cm}

\noindent Emanuela Raffinetti\\
Department of Economics and Management, University of Pavia (Italy)\\
\noindent email: emanuela.raffinetti@unipv.it

\vspace{1cm}

\noindent Giuseppe Toscani\\
Department of Mathematics, University of Pavia (Italy)\\
\noindent email: giuseppe.toscani@unipv.it

\newpage

\begin{abstract}
Inequality measures are quantitative measures that take values in the unit interval, with a zero value characterizing perfect equality. Although  originally proposed to measure economic inequalities, they can be applied to several other situations, in which one is interested in the mutual variability between a set of observations, rather than in their deviations from the mean. While unidimensional measures of inequality, such as the Gini index, are widely known and employed, multidimensional measures, such as Lorenz Zonoids, are difficult to interpret and computationally expensive and, for these reasons, are not much well known. To overcome the problem, in this paper we propose a new scaling invariant multidimensional inequality index,  based on the Fourier transform,  which exhibits a number of interesting properties, and whose application to the multidimensional case is rather straightforward to calculate and interpret.  \end{abstract}
\vspace{0.5cm}

\newpage




%

\section{Introduction}\label{sec:intro}
As recently discussed in \cite{Ban,Eli,Eli3}, the challenge of measuring the statistical heterogeneity of  measures arises in many fields of science and engineering, and it is one of the fundamental features of data analysis. In economics and social sciences size, measures of interest are wealth measures, such as those introduced by \cite{BL,Cou,Cow,GudRaf2023,HN}. Specifically, inequality indices quantify the socio-economic divergence of a given wealth measure from the state of perfect equality. In economics, the most used measure of inequality is the Gini index, first proposed by the Italian statistician Corrado Gini more than a century ago \cite{Gini1,Gini2}. However, although it has had an economic origin, the use of the Gini index has not been limited to wealth alone \cite{HR}. 

A related tool for measuring income and wealth  inequality is the Lorenz function and its graphical representation, the Lorenz curve \cite{Lor}, that plots the percentage of total income earned by the various sectors of the  population, ordered by the increasing size of their incomes.
The Lorenz curve is typically represented as a curve in the unit square of opposite vertices in the origin of the axes and the point $(1,1)$,  starting from the origin and ending at the point  $(1,1)$. 
  
The diagonal of the square exiting the origin is the line of perfect equality, representing a situation in which all individuals have the same income. Since the diagonal is the line of perfect equality, we can say that the closer the Lorenz curve is to the diagonal, the more equal is the
 distribution of income. 
 
This idea of \emph{closeness} between the line of perfect equality and the Lorenz curve can be expressed in many ways, each of which gives rise to a possible measure of inequality.
 Thus, starting from the Lorenz curve, several indices of  inequality can be defined, including the Gini index. Various indices were obtained by looking at the maximal distance between the line of perfect equality and the Lorenz curve, either horizontally or vertically, or alternatively, parallel to the other diagonal of the unit square \cite{Eli}.

Among the enormous amount of research illustrating the fields of application of inequality indices, and of the Gini index in particular, very few consider the case of a multidimensional variable. An exception is the stream of works that consider Lorenz Zonoids (see e.g. \cite{KoshMos96}).
While theoretically sound, Lorenz Zonoids are difficult to implement in practice, especially for computational reasons. 

The above limitation  suggests the development of new inequality indices, that can generalise or substitute the Gini index in the multidimensional case retaining as much of its properties as possible, and that can be of simple implementation.

To this aim, in this paper we introduce a new multidimensional inequality index, which is consistently derived from the one-dimensional index considered in \cite{To1}, based on the Fourier transform of a probability distribution. The new index exhibits a number of interesting properties, and, among others, its value for a multivariate Gaussian distribution is explicitly computable, which renders its implementation in real problems immediate. \\ Among these properties, the new index is built to satisfy the \emph{scaling property on components} \cite{HR}, a property which appears essential when trying to recover the value of the inequality index in a multivariate phenomenon composed by different quantities which are measured by different unit of measures. This is a property which is naturally satisfied by all one-dimensional inequality indices, but it is not satisfied by simply extending the definitions of univariate indices to higher dimensions.

As discussed in \cite{To1}, measures of inequality based on the Fourier transform have some clear advantages. On the one hand, it is very simple to calculate the value taken by the index for probability distributions for whose the characteristic function is explicitly available. In the one-dimensional case this holds, among others, for the Poisson distribution and, for probability measures defined on the whole real line $\R$, for the stable laws \cite{Zol}.  On the other hand, when dealing with a discrete probability measure, the use of the Fourier transform makes it possible to develop very fast computational procedures \cite{Au1,Au2}.

In more detail, in the next Section \ref{sec:multi} we will  detail the new  inequality measure and its main properties. In particular, in Section \ref{sec:scaling} we will show how to modify the one-dimensional version introduced in \cite{To1} in order to satisfy the scaling property along each component.  This analysis shows that  the scaling property of the inequality index is closely related to the concept of Mahalanobis distance \cite{Maha}. Explicit evaluations of the value of the new $n$-dimensional inequality index are  presented in Sections \ref{sec:example} and \ref{sec:gaussian}. In particular, the evaluation in Section \ref{sec:gaussian} of the index in correspondence to a multivariate Gaussian distribution shows that its value is proportional to the multidimensional coefficient of variation of Voinov and Nikulin \cite{VN}, establishing an interesting connection between inequality indices and the multidimensional coefficient of variation via the multivariate Gaussian distribution. Further properties of the Fourier-based inequality index are presented in Section \ref{sec:properties}. Last, a possible proposal for the multivariate Pietra and Gini indices is presented in Section \ref{new-gini}.

\section{A multidimensional Fourier transform measure  of inequality}  \label{sec:multi}

For a given $1\le n \in \N$, we denote by $P_s(\R^n)$, $s \ge 2$, the class of all probability measures $F = F(\bx)$ on the Borel subsets of $\R^n$  such that
\[
m_s(F)  = \int_{\R} |\bx|^s dF(\bx) < + \infty,
\]
where, for a given column vector $\bx = (x_1,x_2,\dots,x_n)^T$,   of dimension $n$,  $|\bx|= \sqrt{\bx^T\bx}$ is the modulus of the vector, i.e. the distance of the point $(x_1,x_2,\dots,x_n)$ from the origin of the cartesian axes.  This condition on $s$ ensures that the covariance matrix $\Sigma$ of a random vector ${\bf X}$ with probability measure $F \in P_s(\R^n)$ is well defined. \\ Further, we denote by $ \tilde P_s (\R^n)$ the class of probability measures $F \in P_s(\R^n)$ which possess a mean value vector ${\bf m}= (m_1,m_2, \dots, m_n)^T$, with nonnegative components $m_k$, $k =1,2,\dots, n$, i.e.
\[
{m}_k (F) =  \int_{\R^n } x_k \, dF(\bx) \ge 0, \quad k =1,2,\dots, n, 
\]
and, at the same time, $|\bm| >0$. Last, we denote with $P_s^+(\R^n)$ the subset of probability measures $F \in P_s(\R^n)$ such that $F(\bx) = 0$ if at least one component $x_k \le 0, k =1,2,\dots, n$. \\
In what follows, for a given $F \in P_s(\R^n)$, we will denote by $\ff(\bxi)$, with $\bxi = (\xi_1,\xi_2,\dots,\xi_n)^T$, its Fourier transform, defined by 
\[
\ff(\bxi) =  \int_{\R^n } e^{-i \bxi^T\bx} \, dF(\bx),
\]
and we denote by  $\mathcal F_s^n$ the class of all Fourier transforms of probability measures in  $P_s(\R^n)$.

\subsection{The one-dimensional case}

As shown in \cite{To1},  both the classical Gini and Pietra indices can be expressed in terms of one-dimensional Fourier transform. As a consequence, the quantitative information on the function $F(x)$ expressed by these indices,   is equ\-ivalently expressed in terms of their Fourier  transformed version, acting on $\ff(\xi)$. In other words, by passing to Fourier transform, we do not lose information about the degree of inequality contained in the probability measure $F(x)$.\\
Let us briefly recall how this result has been derived.  \\ For any fixed constant  $a >0$, denote by $F_a(x)$ the Heaviside step function defined by
 \be\label{he}
 F_a(x) : = \left\{
  \begin{array}{cc}
0 &\quad x< a\\
  1 &\quad x \ge a
  \end{array}
 \right.
 \ee
Given a probability measure $F \in P_s^+(\R)$ with mean value $m$, the Gini index can be expressed by the formula
\be\label{Gin}
G(F) = 1 - \frac 1m\int_{\R_+} (1- F(x))^2\, dx.
\ee
Since $F \in P_s^+(\R)$, $F(x) = 0$ for $x \le 0$. Hence, resorting to the definition of the Heaviside step function  $F_0(x)$, we can write
\[
\int_{\R_+} (1 - F(x))^2 \, dx =\int_\R |F_0(x) - F(x)|^2 \, dx.
\]

For any given pair of probability measures $F,G \in P_s(\R)$, Parseval formula  implies
\be\label{Par}
\int_\R |F(x) - G(x)|^2 \, dx = \frac 1{2\pi} \int_\R |\widehat F(\xi) - \widehat G(\xi)|^2 \, d\xi,
\ee
 where $\widehat F$ and $ \widehat G$ are the Fourier transforms of the probability measures $F,G$. Since
 \be\label{trans}
 \widehat F(\xi) - \widehat G(\xi) = \frac{\ff(\xi) -\gg(\xi)}{i\xi},
 \ee
we conclude with the identity
 \be\label{Par2}
\int_\R |F(x) - G(x)|^2 \, dx = \frac 1{2\pi} \int_\R \frac{|\ff(\xi) -\gg(\xi)|^2}{|\xi|^2} \, d\xi.
\ee

Therefore, for any probability measure $F\in P_s^+(\R)$, the Gini index can be expressed in Fourier transform, as follows:
\be\label{Gin-F}
G(F) = 1 - \frac 1{2\pi m} \int_\R \frac{|1 -\ff(\xi)|^2}{|\xi|^2} \, d\xi.
\ee
Expression \fer{Gin-F} clarifies that in one dimension the Fourier expression of the classical Gini index is a function of a certain distance between probability measures $F$ and $G$, namely
\[
d_2(F,G) = \int_\R \frac{|\ff(\xi)-\gg(\xi)|^2}{|\xi|^2} \, d\xi.
\]
This type of metrics have been extensively studied in connection with the convergence to equilibrium of kinetic equations, as interesting alternatives to more classical entropies \cite{GTW,CaTo,ToTo}. Resorting to this analogy, in \cite{To1} new inequality measures were introduced, some of them related to the supremum distance
\be\label{supre}
d_\infty(F,G) = \sup_{\xi\in \R} \frac{|\ff(\xi)-\gg(\xi)|}{|\xi|}.
\ee
In particular, one new inequality index has been shown to have a deep connection with the supremum distance \fer{supre}, including a number of interesting properties. For a given probability measure $F \in \tilde P_s (\R)$ with mean value $m >0$,  let us consider in \fer{supre} the distance of $F$ from the step function $F_m$, as defined by \fer{he}. Then, the identity
\begin{equations}\nonumber
d_\infty(F,F_m) = &\sup_{\xi\in \R} \frac{|\ff(\xi)-e^{-i\xi\, m}|}{|\xi|} 
= \sup_{\xi\in \R} \frac{|\ff(\xi)e^{i\xi\, m}- 1|}{|\xi|},
\end{equations}
coupled with Lagrange theorem, shows that 
\be\label{otti}
d_\infty(F,F_m) \le \sup_{\xi\in \R} \left| m \ff(\xi)-\ff'(\xi)\right|.
\ee
In  \fer{otti},  $\ff'(\xi)= \ff'_\xi(\xi)$ denotes the derivative of the Fourier transform $\ff(\xi)$ with respect to its argument $\xi$. 
The above inequality suggested the introduction of a new Fourier-based inequality index.
This index reads
\be\label{ine-T} 
T(F) = \frac 12 \sup_{\xi \in \R} \left| \ff(\xi) - \frac{\ff'(\xi)}{\ff'(0)} \right|.
\ee

\subsection{The $n$-dimensional Fourier-based index}\label{sec:nd}
The one-dimensional index $T$ can be easily generalized to any dimension $n >1$. 
For a given probability measure $F \in \tilde P_s (\R^n)$, $n>1$,  such that the mean value $|\bf m|>0$, we  introduce on $\mathcal F_s^n$ a multivariate inequality index $T_n(F)$, expressed by the formula
\be\label{ine-Tn} 
T_n(F) = \frac 1{2| \nabla \ff(\bxi = {\bf 0})|} \sup_{\bxi \in \R^n} \left| \nabla \ff(\bxi = {\bf 0}) \ff(\bxi) - \nabla \ff(\bxi) \right|.
\ee
In definition \fer{ine-Tn},  $\nabla \ff(\bxi)$ denotes the gradient  of the scalar function $\ff(\bxi)$. Indeed,  $F \in P_s(\R^n)$ implies that  $\ff(\bxi)$ is  continuously differentiable, at least up to the order two. It is important to remark that  $i \nabla \ff(\bxi = {\bf 0})$ coincides with the mean value $(m_1(F),m_2(F), \dots, m_n(F))^T$, so that $|\nabla \ff(\bxi = {\bf 0})| = |\bm(F)|$.
\\
The connection of the Fourier-based $T_n$-index with Gini and Pietra indices can be easily established by resorting to standard properties of the Fourier transform. Indeed, for a given $F\in \tilde P_s(\R^n)$, $n \ge 1$, of mean value $\bm$ such that $|\bm| >0$, it holds
\be\label{new-def}
T_n(F) = \frac 1{2|\bm|} \sup_{\bxi \in \R^n} \left| \int_{\R^n } (\bx -\bm) e^{-i \bxi^T\bx} \, dF(\bx) \right|.
\ee
Since
\[
 \left| \int_{\R^n } (\bx -\bm) e^{-i \bxi^T\bx} \, dF(\bx) \right| \le  \int_{\R^n } \left| \bx -\bm\right| \, dF(\bx),
\]
expression \fer{new-def}  clarifies that $T_n(F) \le P_n(F)$, where $P_n(F)$,  given by
\be\label{P-n}
P_n(F) = \frac 1{2|\bm|} \int_{\R^n } \left| \bx -\bm\right| \, dF(\bx),
\ee
is a multivariate version of the classical Pietra index, that, for a probability measure $F\in P_s^+(\R)$ reads
\[
P(F) = \frac 1 m\int_m^\infty (x-m)dF(x)
 = \frac 12 \int_{\R_+} |x-m| \, dF(x).
\]
Moreover, since
\[
\int_{\R^n} |\bx-\bm| \, dF(\bx) = \int_{\R^n}\left| \int_{\R^n}(\bx- \by) \,dF(\by) \right|\, dF(\bx),
\]
it follows that  $P_n(F) \le G_n(F)$, where $G_n(F)$,  given by
\be\label{G-n}
G_n(F) = \frac 1{2|\bm|} \int_{\R^n }\int_{\R^n}  \left| \bx -\bm\right| \, dF(\bx)dF(\by),
\ee
is a multivariate version of the classical Gini index. Indeed, for a probability measure $F\in P_s^+(\R)$, in alternative to \fer{Gin}, Gini index can be written as
\[
G(F) = 
 \frac 1{2m} \int_{\R_+} \int_{\R_+}|x-y| \, dF(x)dF(y).
\]
Finally, for any probability measure  $F\in P_s^+(\R)$ we have the chain of inequalities
\be\label{chain}
0\le T_n(F) \le P_n(F) \le G_n(F) \le 1.
\ee
\begin{remark}
The chain of inequalities \fer{chain} allows to understand the main computational difference between the $n$-dimensional versions of the Pietra and Gini indices and the Fourier-based index. In the case of Pietra and Gini, the evaluation of the  index requires the integration over $\R^n$ of the modulus of a vector, that for $n >1$ does not lead to explicit values, if not in particular simple cases. On the contrary, when the evaluation of $T_n$ is concerned, an evaluation of a modulus is present only when evaluating the supremum. 
\end{remark}
Once the relationships with other well-known inequality indices have been pointed out, we can  move on to list the main properties of the new Fourier-based multivariate inequality index $T_n$.\\
Given  $F \in P_s^+(\R^n)$, the index $T_n$ is bounded from above. This directly follows from the chain of inequalities \fer{chain}. Also,
since for any given $F \in P_s^+(\R^n)$ it holds $|\ff(\bxi)| \le \ff({\bf 0}) = 1$, and 
\[
\frac{\partial \ff(\bxi)}{\partial \xi_k} = -i \int_{(\R_+)^n}  x_k e^{-i \bxi^T\bx} \, dF(\bx), \quad k =1,2,\dots, n,
\]
one obtains the bound
\be\label{posi}
\left| \frac{\partial \ff(\bxi)}{\partial \xi_k}\right| \le \int_{(\R_+)^n}  x_k \left|e^{-i \bxi^T\bx}\right| \, dF(\bx) =m_k,
\ee
which implies $|\nabla \ff(\bxi)| \le |\nabla \ff(\bxi ={\bf 0})| = |\bm|$.

Hence, without resorting to the chain of inequalities \fer{chain}, one shows  that the inequality index $T_n(F)$ satisfies the usual bounds
\be\label{bound-n}
0 \le T_n(F) \le 1,
\ee
and $T_n(F) = 0$ if and only if $\ff(\bxi)$ satisfies the differential equations
\[
\frac{\partial \ff(\bxi)}{\partial \xi_k} = \left. \frac{\partial \ff(\bxi)}{\partial \xi_k}\right|_{\bxi = \bf 0} \ff(\bxi), \quad k =1,2,\dots, n,
\]
with $\ff({\bf 0}) = 1$. 

Thus, as in the one-dimensional case,   $T_n(F)$ vanishes  if and only if $\ff(\bxi) = e^{-i\bxi^T\bm}$, namely if $\ff(\bxi)$ is the Fourier transform of a Dirac delta function located in the point $\bx^T =\bm^T(F)$. Note however that, even if the functional $F$ is defined on the whole class $\tilde P_s(\R^n)$,  the upper bound is lost if the probability measure $F \notin  P_s^+(\R^n)$, since in this case  inequality  \fer{posi} is no more valid. 

\subsection{The scaling property}\label{sec:scaling}
In the rest of this paper, for any given random vector $\bX$ described in terms of its probability measure $F$, we will will identify the functionals $R(\cdot)$ as both $R(F)$ and $R(\bX)$.
It is immediate to show that the functional $T_n(\bX)$ is invariant with respect to the scaling (dilation)
\[
\bX \to c\bX, \quad c >0.
\]
However, as one can easily verify by direct inspection, in the multivariate setting the invariance does not hold anymore  under a more general scaling transformation like $A\bx$, where $A$ is a square non-singular matrix, even if $A$ is a diagonal matrix other than a multiple of the identity matrix. The scaling property  only holds  when the matrix $A$ is orthogonal. \\ This limitation represents a substantial problem when trying to recover the value of the inequality index in a multivariate phenomenon composed by different quantities which are measured by different unit of measures in different countries. Indeed, it is not reasonable for the index to vary if there is a change of unit on a component. 

This unpleasant feature can be circumvented by suitably modifying the point $\bxi$ in which the Fourier transform $\ff$ of $F$ is evaluated. \\
Let us first consider the case in which the random vector ${\bf X} = (X_1,\dots, X_n)^T$ has independent components, so that its covariance matrix $\Sigma$ is diagonal, with diagonal ele\-ments $\sigma_k$, $k =1,2\dots,n$  (the variances of the random variable $X_k$). If $\ff(\bxi)$ is the Fourier transform of the multivariate probability measure of $\bf X$, the change $X_k \to a X_k$, with $a$ positive constant, in the $k$-th component of the vector, induces the same change $\xi_k \to a\xi_k$ in the $k$-th component of the vector $\bxi$, together with a change $\sigma_k \to a^2 \sigma_k$ in its variance. Consequently, the quantity $\xi_k/\sqrt{\sigma_k}$ is invariant with respect to the scaling $X_k \to a X_k$.\\
In other words, by setting 
\be\label{bello}
\ff^*(\bxi) = \ff(\bxi^*), \quad \bxi^* = \left(\frac{\xi_1}{\sqrt{\sigma_1}}, \frac{\xi_2}{\sqrt{\sigma_2}}, \dots,\frac{\xi_n}{\sqrt{\sigma_n}}\right)^T,
\ee
the inequality index $T_n$, evaluated on $\ff^*$, is invariant with respect to the change $X_k \to aX_k$.
\\
The general case in which the random vector ${\bf X} = (X_1, X_2, \dots, X_n)^T$ has a positive definite covariance matrix $\Sigma$ of elements $\sigma_{ij}$, $i,j = 1,2,\dots,n$, can be easily reduced to the previous case by a two step modification of the vector $\bxi$. Let $Z$ denote the orthogonal square matrix such that 
\be\label{dia}
Z^T\Sigma Z = \Lambda, 
\ee
where $\Lambda$ is the diagonal square matrix with elements the eigenvalues $\lambda_k$, $k =1,2,\dots,n$ of $\Sigma$, and let us set
\be\label{bello2}
\ff^*(\bxi) = \ff(\bxi^*), \quad \bxi^* = \left(\frac{Z\xi_1}{\sqrt{\lambda_1}}, \frac{Z\xi_2}{\sqrt{\lambda_2}}, \dots,\frac{Z\xi_n}{\sqrt{\lambda_n}}\right)^T.
\ee
Then, as before, one concludes that the inequality index $T_n$, evaluated on $\ff^*$, is invariant with respect to the change $X_k \to aX_k$.\\
Clearly, the passage from $\bxi$ to $\bxi^*$ corresponds to a change of the mean value. It is a simple exercise to verify that, if $| \nabla \ff(\bxi = {\bf 0})|= |\bm|$, 
\be\label{mah}
 \left| \nabla \ff^*(\bxi = {\bf 0})\right| = \left| \bm^T \Sigma^{-1} \bm \right|^{1/2},
\ee
which equals the Mahalanobis distance of the origin from the probability measure $F$ of mean value $\bm$ and covariance matrix $\Sigma$. \\
We recall that, given a probability measure $F \in \tilde P_s (\R^n)$ of mean value $\bm$ and covariance matrix $\Sigma$, the Mahalanobis distance of a point $\bx=(x_1,x_2, \dots,x_n)^T$ from $F$ is expressed by \cite{Maha}
\be\label{def:M}
d_M(\bx;F) = \sqrt{(\bx-\bm)^T\Sigma^{-1}(\bx- \bm)}.
\ee
Also, given two points $\bx^T$ and $\by^T$ in $\R^n$, the Mahalanobis distance between them with respect to $F$ is given by
\be\label{def:M2}
d_M(\bx, \by;F) = \sqrt{(\bx-\by)^T\Sigma^{-1}(\bx- \by)}.
\ee
This establishes a strict link between the property of scaling invariance of the inequality measure $T_n$ and the Mahalonobis distance. \\
 It is interesting to note that, in a recent work \cite{DelGiudice1}, the Mahalonobis distance is introduced as a measure of heterogeneity, intended as the extent to which contributions to the overall effect size are concentrated in a small subset of variables rather than evenly distributed across the whole set. In line with this consideration, \cite{DelGiudice1} proposed two heterogeneity coefficients for the Mahalanobis distance based on the Gini coefficient. A further investigation of these coefficients is also provided in \cite{DelGiudice2}, especially regarding the related limitations.
 However, connections of the Mahalonobis distance with the property of scaling invariance were not discussed.\\
The proof of identity \fer{mah} requires some simple computations. Since $\ff^*(\bxi) = \ff(\bxi^*)$ we obtain, for $i=1,2,\dots, n$
\[
\frac{\partial \ff^*(\bxi)}{\partial \xi_i} = \sum_{k=1}^n \frac{\partial \ff(\bxi^*)}{\partial \xi^*_k}\frac{\partial \xi^*_k}{\partial \xi_i} .
\]
Thus, according to \fer{bello2}, taking $\xi=0$ we get
\be\label{mean*}
m^*_i =\sum_{k=1}^n m_k\frac{z_{ki}}{\sqrt{\lambda_i}} = \frac{(Z^T\bm)_i}{\sqrt{\lambda_i}} = m_{*,i},
\ee
where, for a given vector $\bx =(x_1,x_2, \dots, x_n)^T$ we denoted
\be\label{basso}
\bx_* = \left(\frac{Z^Tx_1}{\sqrt{\lambda_1}}, \frac{Z^Tx_2}{\sqrt{\lambda_2}}, \dots,\frac{Z^Tx_n}{\sqrt{\lambda_n}}\right)^T
\ee
Therefore
\[
|\bm^*|^2 = \sum_{i=1}^n \frac 1{\lambda_i} (Z^T\bm)_i^2 =  \bm^T Z\Lambda^{-1} Z^T\bm.
\]
The conclusion follows considering that, since $Z$ is the orthogonal matrix that diagonalizes $\Sigma$, (cf. \fer{dia}), it holds
\[
 Z\Lambda^{-1} Z^T = \Sigma^{-1}.
\]
Finally, for a given a multivariate probability measure $F \in \tilde P_s (\R^n)$ of mean value $\bm$  and positive-definite covariance matrix $\Sigma$, a Fourier-based inequality measure of $F$ satisfying the scaling property is obtained by evaluating the index $T_n$ defined in \fer{ine-Tn} in correspondence to the Fourier transform $\ff^*(\bxi)$, as defined in \fer{bello2}. \\ To this aim, let us denote by $\mathcal F^*_s$ the class of all Fourier transform $\ff^*(\bxi) =\ff(\bxi^*)$, with $\bxi^*$ defined in \fer{bello2}, and $\ff(\bxi) \in \mathcal F^n_s$. For any given $\ff^* \in \mathcal F^*_s$ a $n$-dimensional scaling invariant inequality index is defined by 
\begin{equations}\label{ine-T*}
&\tau_n(F) =T_n(\bX_*) = \\
&\frac 1{2} \sup_{\bxi \in \R^n} \frac{\left| \nabla \ff^*(\bxi = {\bf 0}) \ff^*(\bxi) - \nabla \ff^*(\bxi) \right|}{| \nabla \ff^*(\bxi = {\bf 0})|}.
\end{equations}

\begin{remark}\label{rem:scal} It is important to remark that, in view of identity \fer{mean*}, that, if $\ff(\xi)$ is the Fourier transform of the random vector $\bf X$, $\ff^*$ is the Fourier transform of the random vector ${\bf X}_*$, as given by formula \fer{basso}. 
For this reason, all the properties listed in Section \ref{sec:nd} are still valid, only recalling that the passage from $\ff$ to $\ff^*$ requires the passage from $\bm$ to $\bm^*$.
\end{remark}

\subsection{A leading example}\label{sec:example}
A first example, that will allow us to  check that the upper bound in \fer{bound-n} is reached,  refers to the evaluation of the value of the index in correspondence to a multivariate random variable $\bf X$ taking only the two va\-lues $\aa= (a_1,a_2, \dots, a_n)^T$ and $\ab= (b_1,b_2,\dots, b_n)^T$ in $\R^n_+$ with probabilities $1-p$ and respectively $p$, where $0<p<1$. 
Clearly, the probability measure $F$ of $\bf X$ belongs to $P_s^+ (\R^n)$. The mean value $\bm$ and the covariance matrix $\Sigma$ of elements $\sigma_{ij}$ are easily found. One has
\[
\bm = (1-p)\aa + p \ab, \quad \sigma_{ij} = p(1-p) (b_i-a_i)(b_j-a_j).
\]
The Fourier transform of the distribution $F$ of $\bf X$ is given by
\be\label{two-points}
\ff(\bxi) = (1-p)e^{-i\bxi^T\aa} + p e^{-i \bxi^T\ab}.
\ee
Consequently
\begin{equations}
\ff^*(\bxi) = &(1-p)e^{-i((\bxi^*)^T\aa} + p e^{-i(\bxi^*)^T\ab}=\\
& (1-p)e^{-i\bxi^T\aa_*} + p e^{-i\bxi^T\ab_*},
\end{equations}
Therefore
\[
\nabla \ff^*(\bxi) = -i \left[ (1-p)\aa_* e^{-i\bxi^T\aa_*} + p\ab_* e^{-i \bxi^T\ab_*} \right] 
\]
and 
\[
\nabla \ff^*(\bxi = {\bf 0}) = -i \left[ (1-p)\aa_*  + p\ab_*  \right],
\]
so that
\begin{equations}\nonumber
\nabla \ff^*(\bxi = {\bf 0}) \ff^*(\bxi) - \nabla \ff^*(\bxi) &= i\,p(1-p) (\ab_* -\aa_*)\cdot \\
&\left[e^{-i\bxi^T\aa_*} - e^{-i \bxi^T\ab_*}\right].
\end{equations}
Finally
\begin{equations}\nonumber
&\tau_n(F) =\\&\frac 1{2|\bm^*|} p(1-p) |\ab_*-\aa_*|  \sup_{\bxi \in \R^n} \left| e^{-i\bxi^T\aa_*} - e^{-i \bxi^T\ab_*}\right| =
 \\
&\frac 1{2|\bm^*|} p(1-p) |\ab_*-\aa_*|  \sup_{\bxi \in \R^n} \left| 1 - e^{-i \bxi^T(\ab_*-\aa_*)}\right| = \\
& \frac 1{|\bm^*|} p(1-p) |\ab_*-\aa_*| ,
\end{equations}
since, for any vector $\bf d$ such that $|{\bf d}|>0$
\[
 \sup_{\bxi \in \R^n} \left| 1 - e^{-i \bxi^T{\bf d}}\right| =  \sup_{\bxi \in \R^n} \sqrt{ 2[1 - \cos (\bxi^T{\bf d})]} = 2.
\]
Hence, recalling that  $|\bm^*|$ coincides with the Mahalanobis distance of the origin from the probability measure $F$, and that $|\ab_*-\aa_*|$ coincides with the Mahalanobis distance of the two points $\aa$ and $\ab$  with respect to $F$, we conclude with  the value
\be\label{two}
\tau_n(F) = \frac{p(1-p)\sqrt{(\ab-\aa)^T\Sigma^{-1}(\ab- \aa)} }{\sqrt{\bm^T\Sigma^{-1}\bm}}.
\ee
This expression has a structure which is fully compatible with the the one-dimensional formula computed in \cite{To1}, that reads
\[
T(F) = \tau_1(F) = \frac{p(1-p) |b-a|}{(1-p)a  + p b}.
\]
Let us now consider the case in which, for a given positive constant $\e\ll 1$, $p =\e$,  the point $\aa ={\bf 0}$,  while $\ab = \bm/\e$ is located far away, but leaving the mean value $\bm$ unchanged. In this case $\tau_n(F) = 1-\e$, a value which, as $\e \to 0$  converges to the upper bound expressed by the value $1$.

\subsection{The case of a multivariate Gaussian distribution}\label{sec:gaussian}
One of the interesting features of the multivariate inequality index \fer{ine-T*} is that it gives rise to explicit computations in correspondence to multivariate probability distributions which possess an explicit expression in Fourier transform. Among others, a case of primary interest is furnished by the multivariate Gaussian distribution. \\ The Fourier transform of the distribution function $N$ of a multivariate Gaussian variable ${\bf X} = (X_1,X_2,\dots, X_n)^T$ in $\R^n$, $n >1$, is given by the expression
\be\label{gauss}
\ff(\bxi) = \exp\left\{-i \bm^T\bxi - \frac 12 \bxi^T\Sigma \, \bxi \right\}, 
\ee
where $\bm$ is the vector of the mean values  and $\Sigma$ is the $n\times  n$ covariance matrix, with elements
$\sigma_{ij}$.\\
Therefore, if $\bxi^*$ is defined as in \fer{bello2}, while $\bm_*$ as in \fer{basso}, we obtain
\begin{equations}\label{gauss*}
\ff(\bxi^*) =& \exp\left\{-i (\bxi^*)^T\bm - \frac 12 (\bxi^*)^T\Sigma \, \bxi^*\right\}= \\
&\exp\left\{-i \bxi^T\bm_* - \frac 12 |\bxi|^2\right\}.
\end{equations}
Indeed
\begin{equations}\nonumber
&(\bxi^*)^T\Sigma \, \bxi^* =\\ &
\left( \frac{\xi_1}{\sqrt{\lambda_1}}, , \dots,\frac{\xi_n}{\sqrt{\lambda_n}}\right)Z^T\Sigma Z \left( \frac{\xi_1}{\sqrt{\lambda_1}}, , \dots,\frac{\xi_n}{\sqrt{\lambda_n}}\right)^T =\\
&\left( \frac{\xi_1}{\sqrt{\lambda_1}}, , \dots,\frac{\xi_n}{\sqrt{\lambda_n}}\right)\Lambda \left( \frac{\xi_1}{\sqrt{\lambda_1}}, , \dots,\frac{\xi_n}{\sqrt{\lambda_n}}\right)^T =\\
& \sum_{k=1}^n \xi_k^2 = |\bxi|^2.
\end{equations}
From expression \fer{gauss*} we obtain
\be\label{nabla-n}
\nabla \ff(\bxi^*) =\left( -i\bm_* -  \bxi\right) \ff(\bxi^*),
\ee
so that $\nabla\ff(\bxi^*={\bf 0}) = -i\bm_*$.
Finally
\begin{equations}\label{sup-totale}
\tau_n(N) =&  \frac 1{2|\bm_*|} \sup_{\bxi \in \R^n} \left|  \bxi \exp\left\{-i \bxi^T\bm_*- \frac 12|\bxi|^2 \right\}\right|=\nonumber\\
& \frac 1{2|\bm_*|} \sup_{\bxi \in \R^n} | \bxi |\exp\left\{ - \frac 12|\bxi|^2 \right\} =\frac1{2\sqrt e |\bm_*|}\nonumber.
\end{equations}
Now, in view of identity \fer{mah}, the scaling invariant inequality index $\tau_n$ of a multivariate Gaussian distribution $N$ of mean $\bm$ such that $|\bm|>0$, and positive-definite covariance matrix $\Sigma$ is
\be\label{T-Gauss}
\tau_n(N) = \frac1{2\sqrt e} \frac 1{\sqrt{\bm^T\Sigma^{-1}\bm}}.
\ee
Hence, the inequality index of a Gaussian distribution is proportional to the multivariate coefficient of variation considered by Voinov and Nikulin in their book \cite{VN}, where it is claimed that a natural generalization of the coefficient of variation could be based on  the Mahalanobis distance. Consequently they defined, for a given multivariate probability measure $F$ with a mean $\bm$ such that $|\bm|>0$, and positive-definite covariance matrix $\Sigma$ the multivariate coefficient of variation in the form
\[
C_{VN}(F) =  \frac 1{\sqrt{\bm^T\Sigma^{-1}\bm}}
\]
It is well-known that, among all other proposals of multivariate coefficient of variations, the definition of Voinov and Nikulin  is the unique which is invariant under change of scale \cite{Aer}.

\section{Main properties of the inequality index}\label{sec:properties}

As its one-dimensional version, we can further show that the inequality index $\tau_n$ satisfies a number of properties.

Let $F,G\in \tilde P_s(\R^n)$ two probability measures such that their Fourier transforms $\ff^*$ and $\gg^*$ belong to $\mathcal F^*_s$. Suppose moreover that they have the same mean value $\bm$ and covariance matrix $\Sigma$. \\ Then, for any given $\nu \in (0,1)$ it holds 
\begin{equations}\nonumber
&\nu \nabla\ff^*(\bxi= 0) +(1-\nu)\nabla\gg^*(\bxi = {\bf 0}) = \\
&\nabla \ff^*(\bxi={\bf 0} ) = \nabla \gg^*(\bxi={\bf 0}) =-i \bm^*,
 \end{equations}
  so that
\begin{equations}\nonumber
&\tau_n (\nu F + (1-\nu)G) =\frac 1{2|\bm^*|} \sup_{\bxi \in \R^n} \bigg| -i \bm(\nu \ff^*(\bxi) +\\
&(1-\nu)\gg^*(\bxi))-\nu \nabla\ff^*(\bxi) -(1-\nu)\nabla\gg^*(\bxi) \bigg|  = \\
&\frac 1{2|\bm|} \sup_{\bxi \in \R^n} \bigg| \nu(-i \bm \ff^*(\bxi) - \nabla\ff^*(\bxi))+\\
&(1-\nu)(-i\bm \gg^*(\bxi)) - \nabla\gg^*(\bxi))\bigg| \le \\
& \nu\, \tau_n(F) + (1-\nu)\tau_n(G).
 \end{equations}
 This shows the convexity of the functional $\tau_n$ on the set of probability measures with the same mean vector and covariance matrix.
 
Another interesting  property characterizing the inequality index $T_n$ is linked to its behavior when evaluated on convolutions. Let $\bf X$ and $\bf Y$  independent multivariate random variables with probability measures in $\tilde P_s(\R^n)$ of mean values $\bm_\bX$ and, respectively, $\bm_\bY$, and suppose, for simplicity, that they are characterized by diagonal covariance matrices, say $\Sigma_\bX$ (respectively $\Sigma_\bY$). Then, since $\bf X$ and   $\bY$ are independent,   the Fourier  transform $\hh(\bxi)$ of the distribution measure of the sum $\bW ={\bf X}+{\bf Y}$ is equal to the product $\ff(\bxi)\gg(\bxi)$. Moreover,
\[
\bm_\bW = \bm_\bX+\bm_\bY, \quad \Sigma_\bW = \Sigma_\bX + \Sigma_\bY.
\]
Therefore 
\begin{equations}\label{bel-conv}
&\hh^*(\bxi) = \hh(\bxi^{**}), \\ &\bxi^{**} = \left(\frac{\xi_1}{\sqrt{\sigma_{\bX,1}+\sigma_{\bY,1}}},  \dots,\frac{\xi_n}{\sqrt{\sigma_{\bX,n}+\sigma_{\bY,n}}}\right)^T.
\end{equations}
We define
\be\label{con2}
\ff^{**}(\bxi) = \ff(\bxi^{**}), \quad \gg^{**}(\bxi) = \gg(\bxi^{**}).
\ee
Then, since $\hh^*(\bxi)= \ff^{**}(\bxi) \gg^{**}(\bxi) $,
\[
\nabla \hh^*(\bxi) = \nabla\ff^{**}(\bxi) \gg^{**}(\bxi) + \nabla\gg^{**}(\bxi) \ff^{**}(\bxi),
\]
with $\nabla \hh^*(\bxi = {\bf 0}) = -i\bm^*_W$, and
\[
|\bm_\bW^*| = \sqrt{(\bm_\bX +\bm_\bY)^T\Sigma^{-1}_\bW (\bm_\bX +\bm_\bY)}.
\]
On the other hand,  $|\ff^{**}(\bxi)| \le |\ff^{**}({\bf 0})|=1$ and  $|\gg^{**}(\bxi)| \le |\gg^{**}({\bf 0})|=1$, while
$\nabla \ff^{**}(\bxi = {\bf 0}) = -i\bm_\bX^{**}$, $\nabla \gg^{**}(\bxi = {\bf 0}) = -i\bm_\bY^{**}$ with
\[
|\bm_\bX^{**}| = \sqrt{\bm_\bX ^T\Sigma^{-1}_\bW \bm_\bX}, \quad  |\bm_\bY^{**}| = \sqrt{\bm_\bY ^T\Sigma^{-1}_\bW \bm_\bY}.
\]
 Hence
\begin{equations}\nonumber
&\tau_n({\bf X}+{ \bf Y}) =  \\
&
\frac 1{2|\bm_W^*|} \sup_{\bxi \in \R^n} \bigg| \nabla \hh^*(\bxi = {\bf 0}) \hh^*(\bxi)- 
 \nabla \hh^*(\bxi) \bigg|  =\\  &\frac 1{2|\bm_W^* |} \sup_{\bxi \in \R^n} \bigg| \gg^{**}(\bxi)( -i\bm_\bX^{**} \ff^{**}(\bxi) - \nabla \ff^{**}(\bxi))+\\
& \ff^{**}(\bxi)( -i\bm_Y^{**}\gg^{**}(\bxi) - \nabla \gg^{**}(\bxi)) \bigg|\le \\
&
 \frac{|\bm_X^{**}|} {|\bm_W^*|} U_n({\bf X})\cdot \sup_{\bxi \in \R^n}|\gg^{**}(\bxi)|+ \\
 & \frac{|\bm_Y^{**}|} {|\bm_W^* |} U_n({\bf Y})\sup_{\bxi \in \R^n}|\ff^{**}(\bxi)| \le \\
& \frac{|\bm_X^{**}|} {|\bm_W^*|} U_n({\bf X}) + \frac{|\bm_Y^{**}|} {|\bm_W^* |} U_n({\bf Y}).
\end{equations}
Clearly we defined, for the random vector $\bX$  
\be\label{U-n}
U_n(\bX) = \frac 1{2|\bm^{**}|}  \sup_{\bxi \in \R^n} \bigg|  -i\bm_\bX^{**} \ff^{**}(\bxi) - \nabla \ff^{**}(\bxi)\bigg|.
\ee
Finally we have
\begin{equations}\label{convo}
\tau_n({\bf X}+ \bf Y) \le  & \sqrt{\frac{\bm_\bX ^T\Sigma^{-1}_\bW \bm_\bX} {(\bm_\bX +\bm_\bY)^T\Sigma^{-1}_\bW (\bm_\bX +\bm_\bY)}} U_n({\bf X})  + \\ &\sqrt{\frac{\bm_\bY ^T\Sigma^{-1}_\bW \bm_\bY} {(\bm_\bX +\bm_\bY)^T\Sigma^{-1}_\bW (\bm_\bX +\bm_\bY)} }U_n({\bf Y}).
\end{equations}
The case in which the covariance matrices are not diagonal can be treated likewise, by resorting to the same argument of section \ref{sec:scaling}. \\
 Although the consequences of the formula \fer{convo} is not clear in general, when applied to particular cases it provides interesting results that clarify  basic properties of the inequality index $\tau_n$. \\
Let us consider the case in which $\bf X$  belongs to $P_s^+(\R^n)$ and $\bf Y$ is a random variable that takes the constant valued vector $\bm_\bY$ with positive components with probability $1$.  In this simple case  $\gg(\bxi) = e^{-i\bxi^T\bm_\bY}$, $\Sigma_Y = 0$ and $\tau_n({\bf Y}) = U_n(\bY)= 0$.
Consequently  $\Sigma_\bW = \Sigma_\bX$, and $U_n(\bX) = \tau_n(\bX)$. Hence we have
\begin{equations}\label{tide}
&\tau_n({\bf X}+ {\bf Y}) \le\\
& \sqrt{\frac{\bm_\bX ^T\Sigma^{-1}_\bX \bm_\bX} {(\bm_\bX +\bm_\bY)^T\Sigma^{-1}_\bX (\bm_\bX +\bm_\bY)}} \tau_n(\bf X) < \tau_n(\bX).
\end{equations}
 Note that the same result holds even if only one component of the $\bf Y$ variable is bigger than zero, while the others are not. The meaning of inequality \fer{tide} is clear. 
Since in this case $\bf X+Y$ is nothing but $\bf X +\bm_\bY$, which corresponds to adding  constant values $m_k$ to $X_k$, for $k=1, 2,\dots, n$, this property asserts that adding a positive constant value to one or more components  to a certain feature of each agent decreases inequality. This is a well-known property of the one-dimensional Gini index, which says that giving a certain (constant) amount of money to each agent in the society decreases wealth inequality. \\
A second interesting application of formula \fer{convo} follows when the two random vectors $\bX$ and $\bY$ are independent copies of the same probability measure, say $F$. In this case $\Sigma_\bW = 2\Sigma_\bX$, so that $\bxi^{**}= \bxi^*/\sqrt 2$, and $\bm_\bX^{**}  = \bm_\bX^{*}\sqrt 2$. Therefore $U_n(\bX) = \tau_n(\bX\sqrt 2) = \tau_n(\bX)$, in reason of the dilation property. Finally, if the two random vectors $\bX$ and $\bY$ are independent copies of the same probability measure we obtain
\be\label{copy}
\tau_n(\bW) \le \frac 12 \tau_n(\bX) + \frac 12 \tau_n(\bY) = \tau_n(\bX). 
\ee
Clearly, the same result holds any time we consider the value of the inequality index $\tau_n$ relative to the sum $\bW= a\bX$ + $b\bY$, with $a,b$ positive constants, and the two random vectors $\bX$ and $\bY$ independent copies of the same probability measure, so that
\[
\tau_n(\bX+\bY) \le  \tau_n(\bX).
\]
\noindent A third important consequence of inequality \fer{convo} is related to the situation in which the random vector $\bY$ represents a noise (of  mean value $\bm_\bY$ such that $|\bm_\bY| >0$) that is present when measuring the inequality index of $\bX$.  Let us assume that the additive noise is represented by a Gaussian random vector of mean $\bm_\bY$ and covariance matrix proportional to $\Sigma_\bX$, $\sigma^2\Sigma_\bX$. 

If this is the case, we obtain
\begin{equations}\label{new-gau}
&\tau_n({\bf X}+ {\bf Y}) \le\\
& \sqrt{\frac{\bm_\bX ^T\Sigma^{-1}_\bX \bm_\bX} {(\bm_\bX +\bm_\bY)^T\Sigma^{-1}_\bW (\bm_\bX +\bm_\bY)}} \tau_n(\bf X) \\ + 
&\sqrt{\frac{\bm_\bY ^T\Sigma^{-1}_\bY \bm_\bX} {(\bm_\bX +\bm_\bY)^T\Sigma^{-1}_\bW (\bm_\bX +\bm_\bY)}} \tau_n(\bf Y).
\end{equations}
However, considering that $\bY$ is a Gaussian random vector of mean $\bm$ and covariance matrix $\sigma^2\Sigma_\bX$, its inequality index is given by formula \fer{T-Gauss}, that is
\[
\tau_n(\bY) =  \frac1{2\sqrt e} \frac1{\sqrt{\bm^T\Sigma_\bY^{-1}\bm}}.
\]
By substituting this value inside formula \fer{new-gau} we obtain
\begin{equations}\label{new-gau}
&\tau_n({\bf X}+ {\bf Y}) \le\\
& \sqrt{\frac{\bm_\bX ^T\Sigma^{-1}_\bX \bm_\bX} {(\bm_\bX +\bm_\bY)^T\Sigma^{-1}_\bW (\bm_\bX +\bm_\bY)}} \tau_n(\bf X) \\ + 
&\frac 1{\sqrt{{(\bm_\bX +\bm_\bY)^T\Sigma^{-1}_\bW (\bm_\bX +\bm_\bY)}}} \frac 1{2\sqrt e} .
\end{equations}
namely, considering that $\Sigma_\bW = (1+\sigma^2)\Sigma_\bX$, we obtain an explicit upper bound for the multivariate inequality index in terms of the mean value $\bm_\bY$ and  the intensity $\sigma$  of the Gaussian noise.

\begin{remark} It is important to outline that inequality \fer{new-gau} remains valid even if the mean vector of the multivariate Gaussian noise is assumed equal to zero. In this case, by letting $\bm \to 0$ in \fer{new-gau} we obtain the upper bound
\[
\tau_n(\bX+\bY) \le  \sqrt{1+\sigma^2} \left( \tau_n(\bX) + \frac{1}{ |\bm_\bX^*|\sqrt e} \right) .
\] 

\end{remark}

\section{A Proposal for other multivariate indices}\label{new-gini}

As shown in \cite{To1}, as far as univariate indices are considered, the values of the Fourier-based index in correspondence to a random variable taking only the two (positive) values $a$ and $b$ with probabilities $1-p$ and $p$, where $0<p<1$ is equal to the value assumed by the Pietra and Gini one-dimensional inequality indices. Indeed \cite{To1}, if the random variable $X$ is characterized by the Fourier transformed density
\[
\ff(\xi )= (1-p) e^{-i\xi a} + p e^{-i\xi b}, 
\]
it holds 
\[
T(X) = P(X) = G(X).
\]
If we want to maintain the same property for random vectors with $n$ components, it is enough to define  multivariate version of the Pietra and Gini indices, as defined in \fer{P-n} and \fer{G-n}  in  which the Euclidean distance is substituted by the Mahalanobis distance. Given a random vector $\bX$ with probability measure in $P_s(\R^n)$, with mean $\bm$ and covariance matrix $\Sigma$, this modification induces the following definitions for the multivariate inequality indices of Pietra and Gini
\be\label{P-nM}
\pi_n(\bX) = \frac 12  \int_{\R^n} \sqrt{\frac{(\bx -  \bm)^T\Sigma^{-1}(\bx- \bm)}
{\bm^T\Sigma^{-1} \bm}}dF(\bx),
\ee
and
\be
\label{G-nM}
\gamma_n(\bX) = 
\frac 12  \int_{\R^n} \sqrt{\frac{(\bx -  \by)^T\Sigma^{-1}( \bx-\by)}{\bm^T\Sigma^{-1} \bm}}dF(\bx)\, dF(\by).
\ee
These inequality indices satisfy the scaling property and constitute a reasonable generalization to a multivariate setting of the univariate indices.

It is interesting to remark that also these indices, when evaluated in correspondence to a Gaussian density, give a value proportional to the coefficient of variation of Voinov and Nikulin, with  constants $C_\pi$  (respectively $C_\gamma$  such that $C_\pi <C_\gamma$, and $C_\pi$ bigger than the proportionality constant in \fer{T-Gauss}. 

\section{Conclusions}
In this work, we introduced and studied a new index of inequality for multivariate probability measures, defined in terms of the Fourier transform. In particular,  the new index is built to satisfy the \emph{scaling property on components} \cite{HR}, a property which is necessary  in a multivariate phenomenon to ensure a constant value in the presence of a unit change on one or more components. The search for this requirement uncovered an interesting connection between the scaling property and the Mahalanobis distance, that  suggests a new way to define both Gini and Pietra indices in a multivariate setting. Further studies on this issue are in progress.

%
%
%
%

\section*{Acknowledgments}
The paper is the result of a close collaboration between the three authors. G.T. acknowledges partial support of  IMATI (Institute of Applied  Mathematics and Information Technologies ``Enrico Magenes''.

\end{document}